\documentclass{aastex63}

\usepackage{amsmath}

\begin{document}

\title{The averaged broadband spectral energy distribution study of Fermi bright BL Lac objects}

\correspondingauthor{Rui Xue}
\email{ruixue@zjnu.edu.cn}
\correspondingauthor{Junhui Fan}
\email{fjh@gzhu.edu.cn}

\author[0000-0001-8244-1229]{Hubing Xiao}
\affiliation{Shanghai Key Lab for Astrophysics, Shanghai Normal University, Shanghai 200234, China}

\author[0000-0002-8626-8686]{Haitao Cao}
\affiliation{School of Information Engineering, Guangzhou Panyu Polytechnic, Guangzhou 511483, China}

\author[0000-0003-1721-151X]{Rui Xue}
\affiliation{Department of Physics, Zhejiang Normal University, Jinhua 321004, China}

\author{Zhihao Ouyang}
\affiliation{Shanghai Key Lab for Astrophysics, Shanghai Normal University, Shanghai 200234, China}

\author[0000-0001-8485-2814]{Shaohua Zhang}
\affiliation{Shanghai Key Lab for Astrophysics, Shanghai Normal University, Shanghai 200234, China}

\author[0009-0001-7397-1727]{Junping Chen}
\affiliation{School of Physics and Astronomy, Sun Yat-Sen University, Zhuhai 519082, China}

\author{Zhijian Luo}
\affiliation{Shanghai Key Lab for Astrophysics, Shanghai Normal University, Shanghai 200234, China}

\author{Jianghe Yang}
\affiliation{Department of Physics and Electronics Science, Hunan University of Arts and Science, Changde 415000, China}

\author[0000-0002-5929-0968]{Junhui Fan}
\affiliation{Center for Astrophysics, Guangzhou University, Guangzhou 510006, China}
\affiliation{Great Bay Brand Center of the National Astronomical Data Center, Guangzhou 510006, China}
\affiliation{Key Laboratory for Astronomical Observation and Technology of Guangzhou, Guangzhou 510006, China}
\affiliation{Astronomy Science and Technology Research Laboratory of Department of Education of Guangdong Province, Guangzhou 510006, China}

\begin{abstract}
The physics-determined broadband spectral energy distributions (SEDs) of blazars have been widely used to study the property during their flaring/outburst states, while the non-flaring state takes up most of their lifetime and the general property of blazars has been barely discussed.
In this work, for the first time, we used the archival data and employed the physics-determined SED processing method to form approximately average-state SEDs for 513 \textit{Fermi} bright BL Lacs.
In general, we found that the magnetic field ($B$) is weaker than those obtained for flaring/outburst state by nearly one order of magnitude, and the dissipation region size ($R$) is larger than those obtained for flaring/outburst state, suggesting that the dissipation region could be more extend and less magnetized.
A correlation between the synchrotron-self Compton (SSC) peak frequency ($\log \nu_{\rm ssc}$) against the synchrotron peak frequency ($\log \nu_{\rm sy}$) suggest that the inverse Compton scattering of HBLs suffer a significant Klein-Nishina (KN) suppression, we quantified the condition of KN suppression by determining the critical synchrotron peak frequency ($\nu_{\rm sy}^{\rm c}$) and found 359 out of 513 sources in our sample suffer KN suppression.
Furthermore, our analysis of the relationship between synchrotron curvature ($1/b_{\rm sy}$) and $\log \nu_{\rm sy}$ indicates that the energy-dependent probability acceleration (EDPA) mechanism may dominate the particle acceleration in BL Lac jets.
\end{abstract}


\section{Introduction}
\label{1}
Blazars, the most extreme subclasses of active galactic nuclei (AGNs), show extreme observation properties including rapid and strong variability, high and variable polarization, strong and variable $\gamma$-ray emission and apparent superluminal motions \citep{Wills1992, Urry1995, Fan2002, Fan2004, Rani2013, Fan2014, Lyutikov2017, Xiao2019, Xiao2022MNRAS}.
Based on their optical continuum, blazars are categorized into two subclasses: BL Lacertae objects (BL Lacs) and flat spectrum radio quasars (FSRQs). 
BL Lacs are characterized by spectra with no or weak emission lines (rest-frame equivalent width, $\rm EW < 5 \AA$), whereas FSRQs display strong emission line features ($\rm EW \ge 5 \AA$) \citep{Urry1995, Scarpa1997}.

The emission from a blazar is dominated by a relativistic jet that points towards the observer, characterized by a two-bump broadband spectral energy distribution (SED) in a ${\rm log} \nu$-${\rm log} \nu F_{\nu}$ diagram. 
The low-energy bump, which peaks between the millimeter and soft X-ray bands, is attributed to synchrotron emission from relativistic electrons. 
The high-energy bump, peaking in the MeV to GeV range, is primarily due to inverse Compton (IC) scattering in the leptonic scenario. 
The IC process can be further classified as synchrotron self-Compton (SSC) if the soft seed photons originate from the synchrotron radiation of the same electron population, or as external Compton (EC) if the seed photons come from external regions such as the accretion disk \citep{Dermer1993}, the broad-line region (BLR) \citep{Sikora1994, Fan2006}, or the dusty torus (DT) \citep{Blazejowski2000, Arbeiter2002, Sokolov2005}.
Additionally, the hadronic model can also explain the high-energy bump through proton synchrotron radiation and emission from secondary cascades \citep{Mucke2001, Dimitrakoudis2012, Zheng2013, Diltz2015, Xue2022PRD106, Xue2023PRD107, Wang2024ApJS271}, and also employed to explain variability patterns \citep{Mastichiadis2013MNRAS434}.
And the hadronic model has gained more attention after the detection of neutrino association with blazar flare activity, such as that observed in TXS 0506+056 \citep{IceCube2018, Xue2019ApJ886, Xue2021ApJ906}.

The study of blazar SEDs has been performed, via two different methods, in many previous works \citep{Landau1986, Abdo2010ApJ716, Fan2016ApJS, Krauss2016A&A591, Tan2020, Yang2022ApJS, Chen2023ApJS268, Xiao2024RAA24, Hu2024MNRAS}.
In the first kind, namely statistic-determined broadband SED, the SED is phenomenologically structured by a log-parabolic (or polynomial) function and the the goodness of a fit is mainly evaluated statistically (e.g., $\chi^{2}$ statistic).
The advantage of statistic-determined SED fitting is obvious, it is relatively straightforward and can be flexibly applied to large samples using non-simultaneously archival data.
\citet{Giommi1995A&AS} studied multi-band spectra for a large number of radio-selected and X-ray-selected BL Lac objects (RBLs and XBLs, respectively).
They found systematic differences between the spectra of RBLs and XBLs: RBLs exhibited synchrotron peaks at millimeter/infrared (IR) bands, while XBLs exhibited synchrotron peaks at ultraviolet (UV)/soft X-ray bands.
Further study demonstrated a blazar sequence, indicating that blazars with larger bolometric luminosity, primarily flat-spectrum radio quasars (FSRQs), tend to have `redder' SEDs and lower synchrotron peak frequencies ($\nu_{\rm sy}$). 
Conversely, blazars with smaller bolometric luminosity, mainly BL Lacs, tend to have `bluer' SEDs and higher synchrotron peak frequencies \citep{Fossati1998, Ghisellini2008MNRAS387}.
Subsequent studies have classified blazar into subgroups, namely low synchrotron-peaked blazars (LSPs), intermediate synchrotron-peaked blazars (ISPs) and high synchrotron-peaked blazars (HSPs), based on the derived $\log \nu_{\rm sy}$ \citep[e.g.,][]{Abdo2010ApJ716, Fan2016ApJS, Yang2022ApJS}.
Alternatively, blazar classification based on the inverse Compton (IC) peak frequency is also explored \citep{Yang2023SCPMA}, and the classification is also related to the energy distribution of relativistic particles \citep{Xiao2024ApJ966}.
While, we should keep in mind that the location of the synchrotron or IC peak frequency could shift significantly different from flaring state to quiescent state, and the shifts significantly weaken the robustness of classifications based solely on synchrotron/IC peak frequencies (\citealp[Mrk 501,][]{Acciari2011ApJ729}; \citealp[1ES 1959+650,][]{Acciari2020A&A638}; \citealp[VER J0521+211,][]{Adams2022ApJ932}).
However, an unignorable disadvantage of statistic-determined SED fitting is that the property of dissipation region, e.g., magnetic field strength ($B$), dissipation region size ($R_{\rm diss}$), and particle energy distribution cannot be directly constrained, these parameters are keys to understand the detail of blazar emission and particle acceleration mechanism in the jets.
This shortage can be solved using the second kind of method, the physics-determined broadband SED fitting. 
In the conventional standard `one-zone' model, the observed quantities are related to the intrinsic physical properties of the dissipation region in the jets \citep[e.g.,][]{Tavecchio1998, Ghisellini2009MNRAS397}.
While, one should notice that there are caveats of the physics-determined broadband SED fitting methods.
The physics-determined broadband SED fitting results are typically done by eye without a statistical evaluation of the best-fit parameters, and therefore does not allow to determine uncertainties.
The degeneracy between $B$ and $\delta$ needs to be carefully addressed, for instance, constrain one of them in a reasonable range.
\citet{Abdo2011ApJ736} studied the broadband SED during the $\gamma$-ray activity for Mrk 421 with a leptonic SSC model, and obtained that the magnetic field is $B = 3.8 \times 10^{-2} \ \rm Gs$, the Doppler factor is $\delta = 21$, dissipation region radius is $R = 5.2 \times 10^{16} \ \rm cm$, the minimum electron Lorentz factor is $\gamma_{\rm min} = 8 \times 10^{2}$, the high-energy electron spectral index is $p_{\rm 3} = 4.7$. 
These parameter values have also been applied to the SED modeling of Mrk 421 in other works \citep[e.g.,][]{Aleksic2015}.
Besides, in previous works, people also attempted physical modeling use quasi-simultaneous data for a large amount of blazars to explore blazar emission properties \citep[e.g.,][]{Tan2020, Chen2023ApJS268}.
Recently, we performed the broadband physical SED studies for 5 TeV FSRQs and found that during the TeV activity the external photons for IC scattering comes from the DT \citep{Xiao2024RAA24}.
However, the constraints on the physical properties of the dissipation region are typically obtained for only a few blazars, which are simultaneously observed with multiple instruments covering most of the broadband emission.
And the simultaneous observations are usually triggered by special events, such as flares or outbursts, which are only snapshots of blazars.
Recent study of flare duty-cycle suggested that most of the blazars show flare duty cycle less than 0.1 \citep{Yoshida2023ApJ954}, therefore, the physical property of blazar obtained via simultaneously observed data is strongly biased by the flare or outburst and can hardly represent the normal property of blazars.

To understand general physical property of blazar jets, in the current work, we construct average-state (including data from both flaring and non-flaring states) physics-determined SEDs in the frame of leptonic model using available historical data.
And, in this work, we focus our study exclusively on BL Lac objects, as their soft photon fields have a relatively clean and well-understood origin. 
Specifically, the seed photons for the IC scattering in BL Lacs are generally believed to originate from synchrotron radiation produced by the same population of electron, making the SSC process sufficient to explain the high-energy hump in their SEDs.
In contrast, FSRQs have more complex soft photon environments, with possible contributions from the accretion disk, BLR and DT. 
These multiple photon sources introduce significant uncertainties in modeling the high-energy component of their SEDs, and these uncertainties can hardly be removed when we build the broadband SED based on the unfiltered historical data.
This paper is organized as follows.
In Section \ref{sec: sam}, we introduce our sample and the SED modelling method; 
the results and related discussion are presented in Section \ref{sec: res}; 
and conclusions are presented in Section \ref{sec: con}.

\section{Sample and Methodology}
\label{sec: sam}
We search multi-wavelength observations for the BL Lacs in the latest \textit{Fermi} $\gamma$-ray source catalog, Fermi Large Area Telescope Fourth Source Catalog Data Release 4, \citep[4FGL\_DR4,][]{Abdollahi2022ApJS260, Ballet2024} in the Space Science Data Center (SSDC) Sky Explorer \footnote{https://tools.ssdc.asi.it/}.
In order to construct the characteristic two-hump structured SEDs, we build a sample of BL Lacs that with available all four bands (radio, optical, X-ray, and $\gamma$-ray) archival data in SSDC.
In total, we managed to build a sample with 513 BL Lacs, and list them in Table \ref{SED_shape}.

During the SED modeling, we employed a steady state electron energy distribution of the log-parabolic-power-law (LPPL) model, which is expressed as follows
\begin{equation}
N({\gamma})=\left\{
\begin{array}{llr}
N_{\rm 0} (\gamma / \gamma_{\rm 0})^{-s}    \ \ \   ,\gamma \leq \gamma_{\rm 0} \\
\\
N_{\rm 0} (\gamma / \gamma_{\rm 0})^{-(s+r \cdot \log (\gamma / \gamma_{\rm 0}))}    \ \ \    ,\gamma > \gamma_{\rm 0}, 
\end{array} \right.
\label{particle_dis_eq}
\end{equation}
where $N$ is the number density of particles at the energy of Lorentz factor, $\gamma$, $N_{\rm 0}$ and $\gamma_{\rm 0}$ are the normalization parameter and the turnover energy, $s$ is the spectral index, and $r$ is the spectral curvature \citep{Massaro2004a, Massaro2004b, Massaro2006, Tramacere2011}. 
This model describes particle energy distribution of a power law function at the low energy head then becomes a log-parabola function at its high energy tail, it has been used by authors \citep{Tramacere2009, Tramacere2011} and will be also employed throughout this paper.
We employ the \textit{Jets SED modeller and fitting Tool} (\textit{JetSet, version 1.3.0}) \footnote{https://jetset.readthedocs.io/en/latest/index.html}, which is a scientific tool for building and analyzing jet models, \citep{Tramacere2009, Tramacere2011, Tramacere2020} to build broadband SEDs for the sources in our sample.
We employ a semi-manual SED fitting method to construct broadband spectral energy distributions (SEDs) for the sources in our sample. The fitting process consists of two main steps.
First, we use the JetSet to fit the overall shape of the SED and to obtain initial estimates of the physical parameters.
We used the `ObsConstrain' and `constrain\_SSC\_model' functionalities to construct a general SED shape and to obtain initial estimates for key physical parameters such as the magnetic field strength ($B$), Doppler factor ($\delta$), and electron energy distribution parameters.
Second, these initial values are manually fine-tuned to avoid misrepresenting the SED due to the presence of synchrotron self-absorption in the radio band or contamination from flare-state observations. 
This approach allows us to better approximate the average-state SED for each source.

\section{Results and discussions}\label{sec: res}
\subsection{The property of this sample and SED fitting}
\begin{figure}[htbp]
\centering
\includegraphics[scale=0.7]{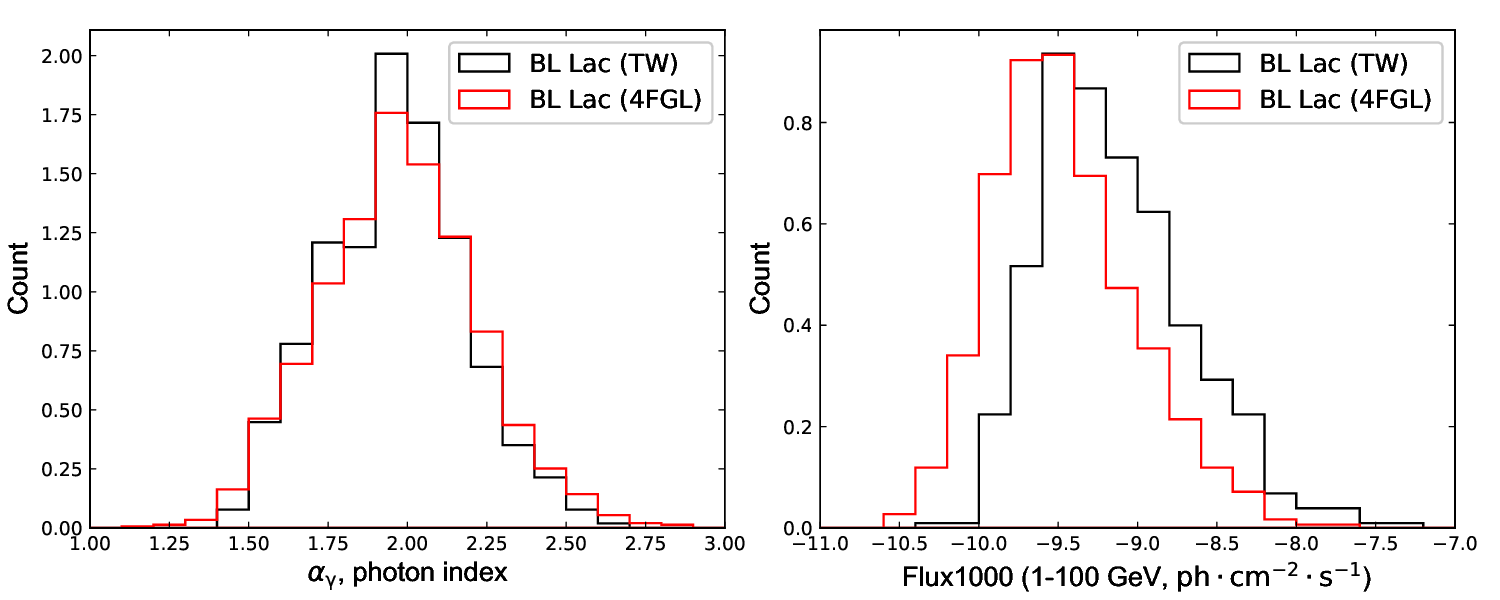}
\caption{The comparison between the BL Lacs in this work and those in 4FGL\_DR4. 
There are 20 bins for each distribution of parameters, and the distributions are normalized to one.}
\label{af_hist}
\end{figure}

In our sample, the 513 BL Lacs have redshifts ranging from 0.01 to 2.284.
A comparison between the 513 BL Lacs in our sample and those 1490 BL Lacs in 4FGL\_DR4 regarding two aspects: the $\gamma$-ray photon index ($\alpha_{\gamma}$) and the integral photon flux (Flux1000) in the $\gamma$-ray 1-100 GeV band.
The comparison is illustrated in Figure \ref{af_hist}.
In the left panel of Figure \ref{af_hist}, the $\alpha_{\gamma}$ shares a similar distribution between the BL Lacs in this work and those in 4FGL\_DR4, with the Kolmogorov–Smirnov (KS) test giving a probability of 0.37. 
In the right panel of Figure \ref{af_hist}, the Flux1000 shows different distributions between the BL Lacs in this work and those in 4FGL\_DR4, with the KS test giving a probability of $3.7 \times 10^{-25}$.
The results of the distributions and the KS tests suggest that the BL Lacs in our sample are averagely brighter than the BL Lacs in 4FGL\_DR4, indicating that we are using a brighter sample of Fermi BL Lacs in this work.
There are two reasons for us to collect this brighter \textit{Fermi} BL Lac sample unintentionally. 
On one hand, we try to collect BL Lacs with more available multi-wavelength data; on the other hand, brighter sources are more likely to be observed in multi-bands.

\begin{figure}
\centering
\includegraphics[width=250pt]{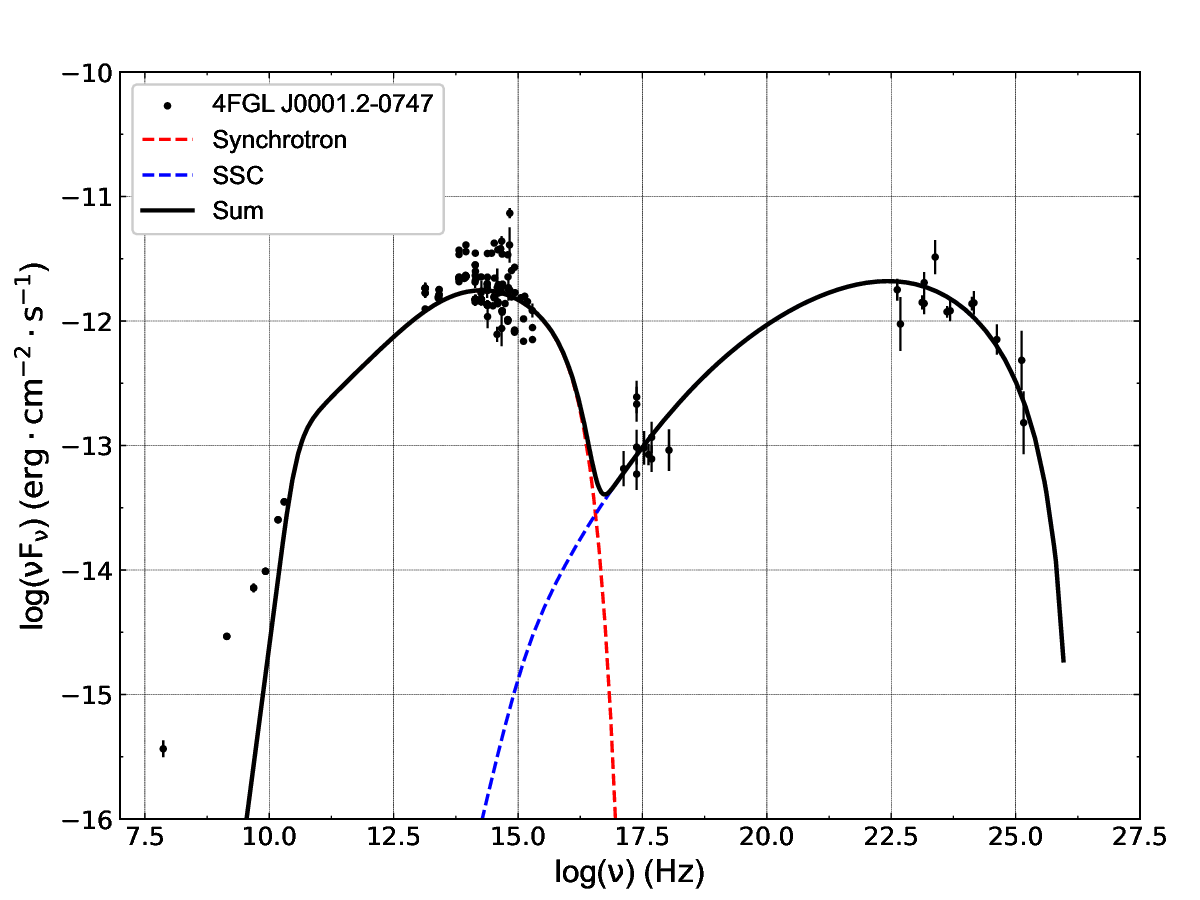}
\includegraphics[width=250pt]{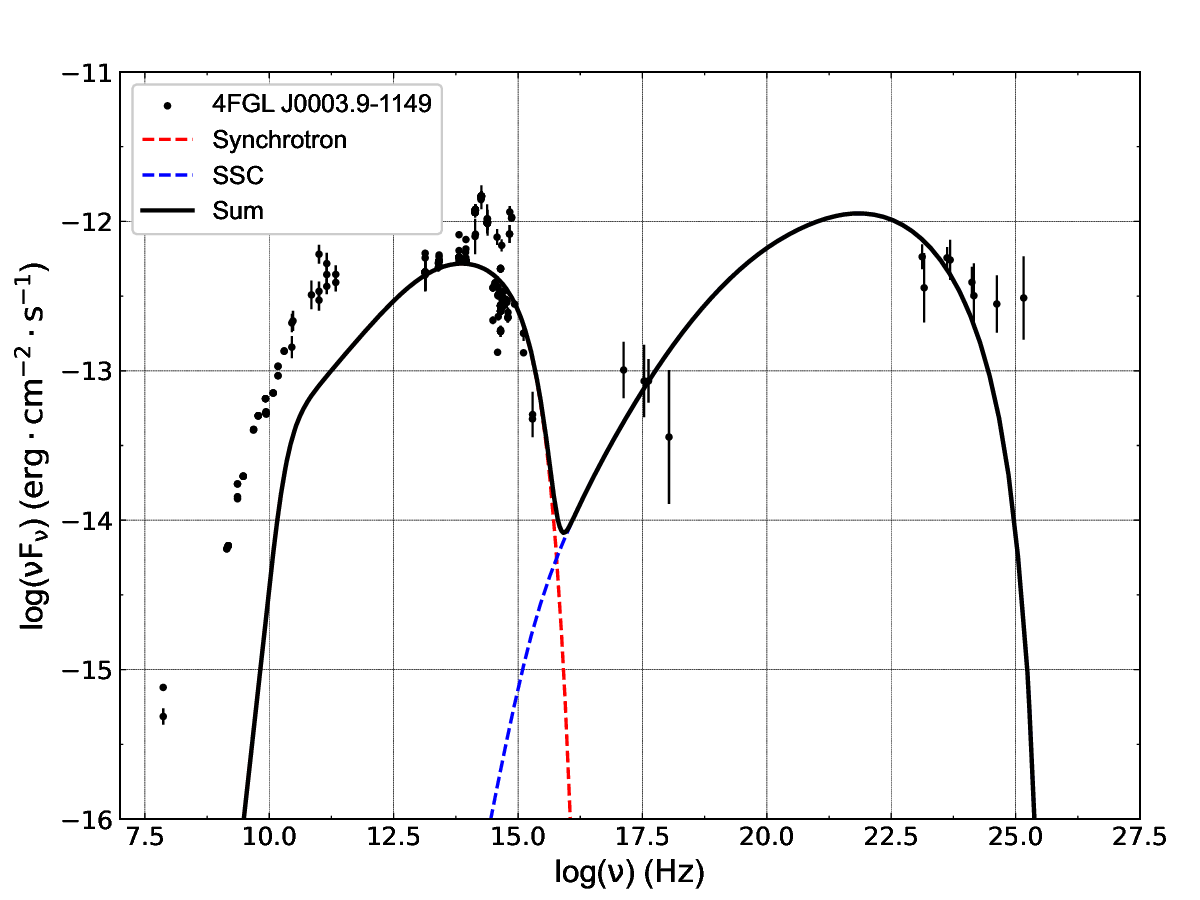}
\caption{The SED modelling plots for 513 Fermi bright BL Lacs.
The dashed-red curve represents the synchrotron component, the dashed-blue curve the SSC component, and the solid-black line represents the summed spectrum.
Only 2 plots (4FGL J0001.2-0747 and 4FGL J0003.9-1149) are shown here, and the remaining plots are available in the online journal.}
\label{SED_plot}
\end{figure}

We managed to fit the broadband SEDs for the 513 Fermi bright BL Lacs using the SSC model through \textit{JetSet}, and demonstrated SED plots in Figure \ref{SED_plot}.

\subsection{The shape parameters of SED}
We employ \textit{JetSet} to complete physics-determined SED fitting for the BL Lacs in our sample.
And, we derive the broadband average-state SEDs along with the basic shape parameters, including the peak frequencies ($\log \nu_{\rm sy}$ and $\log \nu_{\rm ssc}$) and their corresponding peak fluxes ($\log (\nu_{\rm sy}f_{\rm sy})$ and $\log (\nu_{\rm ssc}f_{\rm ssc})$) for the two components.
We summarize these shape parameters — namely, the synchrotron peak and SSC peak measurements — in Table \ref{SED_shape}.

\begin{table}[htbp]
\centering
\caption{The SED shape parameters}
\label{SED_shape}
\begin{tabular}{lccccccc}
\hline
4FGL name   &  z & $\log \nu_{\rm sy}$ &  $\log (\nu_{\rm sy}f_{\rm sy})$ & $\log \nu_{\rm ssc}$ & $\log (\nu_{\rm ssc}f_{\rm ssc})$ & Class & Reduced-$\chi^{2}$ \\
  &  &  Hz & ${\rm erg \cdot cm^{-2} \cdot s^{-1}}$ & Hz & ${\rm erg \cdot cm^{-2} \cdot s^{-1}}$ &  &  \\
(1) & (2) & (3) & (4) & (5) & (6) & (7) & (8) \\
\hline
J0001.2-0747 &	0.24	&	14.27	&	-11.75	&	22.38	&	-11.68	& IBL	& 0.45	 \\
J0003.9-1149 &	0.86	&	13.89	&	-12.28	&	21.89	&	-11.95	& IBL	& 10.39	 \\
J0006.3-0620 &	0.347	&	13.24	&	-11.22	&	20.75	&	-11.33	& LBL	& 545.32 \\
J0009.1+0628 &	1.56	&	13.97	&	-11.88	&	22.12	&	-11.64	& IBL	& 0.34	 \\
J0009.8-4317 &	0.56	&	16.02	&	-12.13	&	23.83	&	-12.35	& HBL	& 0.69	 \\
...          &    ...   &    ...    &     ...   &    ...    &     ...   & ...   &  ...  \\
\hline
\end{tabular}
\tablecomments{
Column (1) 4FGL name;
column (2) redshift;
column (3) synchrotron peak frequency in logarithm, $\nu$ in units of Hz;
column (4) synchrotron peak flux in logarithm, $f$ in units of $\rm {erg/cm^{2}/s}$;
column (5) SSC peak frequency in logarithm;
column (6) SSC peak flux in logarithm;
column (7) classification based on the criterion of \citet{Yang2022ApJS}, `LBL' stands for the lower synchrotron-peaked ($\log \nu_{\rm sy} {\rm (Hz)} \leq 13.7$) BL Lac, `IBL' stands for the intermediate synchrotron-peaked ($13.7 < \log \nu_{\rm sy} {\rm (Hz)} \leq 14.9$) blazars and `HBL' stands for the higher synchrotron-peaked ($\log \nu_{\rm sy} {\rm (Hz)} > 14.9$) BL Lac;
column (8) the value of reduced $\chi^{2}$.
Only five items are displayed, the entire table is available in machine-readable form.}
\end{table}

In this work, we construct physics-determined SEDs and derive corresponding measurements for a sample of bright \textit{Fermi} BL Lac objects.
These sources were also included in previous studies, where they were analyzed using statistical-determined SEDs.
Specifically, we compare the synchrotron peak measurements obtained in this work with those reported in \citet{Yang2022ApJS}, and the IC peak measurements with those presented in \citet{Yang2023SCPMA}.

\begin{figure}[htbp]
\centering
\includegraphics[scale=0.70]{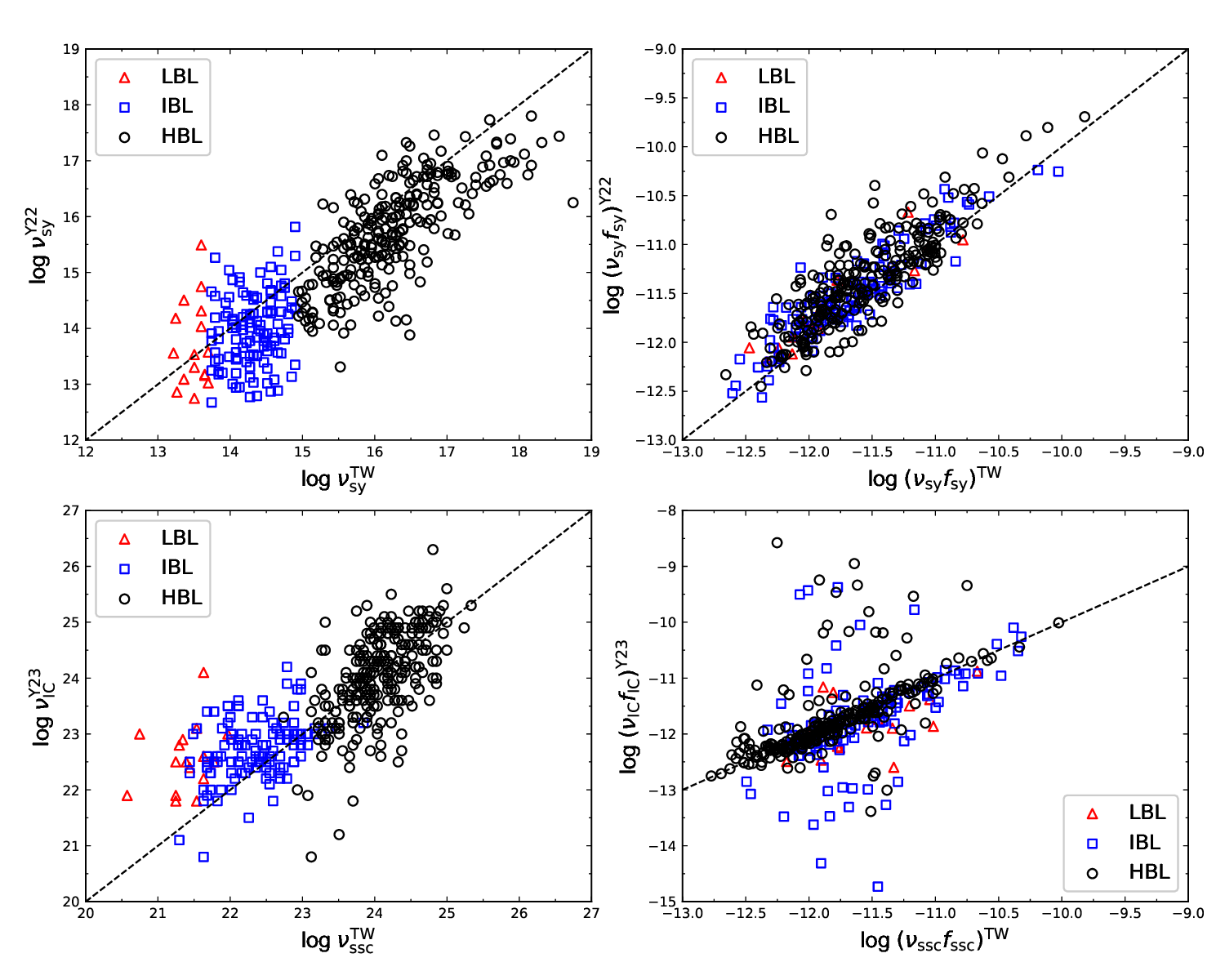}
\caption{The comparison between the shape measurements from this work and that from literature.
The red triangle stands for the LBL, the blue square stands for the IBL and the black circle stands for the HBL.
The top panel shows the comparison of the synchrotron measurements from this work and those from \citet{Yang2022ApJS}; 
the bottom panels show the comparison of the IC measurements from this work and those from \citet{Yang2023SCPMA}.}
\label{vf_com}
\end{figure}

The comparison of $\log \nu_{\rm sy}$ and $\log (\nu_{\rm sy}f_{\rm sy})$ between our results and those of \citet{Yang2022ApJS}, in the top panel of Figure \ref{vf_com}, indicates that we have slightly larger $\log \nu_{\rm sy}$ and smaller $\log (\nu_{\rm sy}f_{\rm sy})$ for these BL Lacs than Yang.
The comparison of the IC measurements between this work and \citet{Yang2023SCPMA} is presented in the bottom panel.
We find that the BL Lacs in our sample exhibit systematically lower values of $\log \nu_{\rm IC}$ compared to theirs.
Additionally, there is a larger scatter in the comparison of $\log (\nu_{\rm IC}f_{\rm IC})$, with our results generally showing higher $\log (\nu_{\rm IC}f_{\rm IC})$ values for most BL Lacs.

The shape measurements derived from the physics-determined SED method exhibit noticeable differences compared to those obtained using the statistic-determined SED approach.
Through our comparisons, we find that the synchrotron peak tends to shift toward higher energy bands in the physics-determined SEDs, which may introduce uncertainties in $\log \nu_{\rm sy}$-based classifications, particularly for IBLs.
Moreover, the synchrotron flux appears to be larger in the statistic-determined SEDs.
A larger value of $\log (\nu_{\rm sy}f_{\rm sy})$ could attribute to the optical enhancement caused by contamination from the accretion disk thermal emission.

\subsection{The physical parameters of SED}
There are nine physical parameters derived from the broadband averaged-SED fitting. 
These include three parameters that describe the dissipation region properties: the dissipation region size ($R$), the magnetic field strength ($B$), and the Doppler factor ($\delta$). 
And, there are six parameters that describe the electron energy distribution (EED): the number density of particles ($N$), the lower energy cutoff of the electrons ($\gamma_{\rm min}$), the turnover energy of the electrons ($\gamma_{\rm 0}$), the high energy cutoff of the electrons ($\gamma_{\rm max}$), the electron spectral index ($s$), and the electron spectral curvature ($r$).
We list these nine physical parameters in Table \ref{SED_physical} and illustrate their distributions in the Appendix A.

\begin{table}[htbp]
\centering
\caption{The SED physical parameters}
\label{SED_physical}
\begin{tabular}{lccccccccc}
\hline
4FGL name & R  &  B & $\delta$ & N & ${\log}\gamma_{\rm min}$ & ${\log}\gamma_{\rm max}$ & ${\log}\gamma_{\rm 0}$ & s & r  \\
    &  cm &  Gs &     & ${\rm cm^{-3}}$ &     &     &     &     &       \\
(1) & (2) & (3) & (4) &    (5)          & (6) & (7) & (8) & (9) & (10)  \\
\hline
J0001.2-0747 &	4.6E+16	 &	0.02	&	22.6 &	50	 &	2.00 & 4.90	&	3.43 &	2.2	&	0.6  \\
J0003.9-1149 &	1E+17	 &	0.01	&	28.6 &	28	 &	2.18 & 4.65	&	3.45 &	2.2	&	0.6  \\
J0006.3-0620 &	2.4E+17	 &	0.01	&	25	 &	4.2	 &	2.36 & 4.54	&	3.41 &	2.2	&	1.3  \\
J0009.1+0628 &	1.85E+17 &	0.018	&	30	 &	18.3 &	2.00 & 4.81	&	3.65 &	2.2	&	1.0  \\
J0009.8-4317 &	2.8E+15	 &	0.6	    &	25	 &	2700 &	1.30 & 4.88	&	3.51 &	2.2	&	0.5  \\
...          &  ...      & ...      & ...    &  ...  &  ...  &  ... &   ...  &  ... &   ...  \\
\hline
\end{tabular}
\tablecomments{
Column (1) 4FGL name;
column (2) dissipation region size, $R$, in units of cm;
column (3) magnetic field strength, $B$, in units of Gs;
column (4) Doppler beaming factor;
column (5) number density, $N$, in units of ${\rm cm^{-3}}$;
column (6) logarithmic low energy cut-off;
column (7) logarithmic turnover energy;
column (8) logarithmic high energy cut-off;
column (9) low energy (LE) spectral slope;
column (10) spectral curvature.
Only five items are displayed, the entire table is available in machine-readable form.
}
\end{table}

For the three dissipation region parameters, the initial estimates of $\delta$ is set to be 20 according to the results of Doppler factor reported in literature \citep{Fan2009PASJ, Hovatta2009, Xiao2022ApJ_1}, and thus the degeneration between $B$ and $\delta$ can be relatively restricted.
At last, the we have $\delta$ ranges from 9 to 37 and $B$ ranges from 0.0015 to 3.2 G, the average of $B$ is significantly smaller than those used in literature by nearly one order of magnitude \citep[e.g.,][]{Tan2020, Chen2023ApJS268, Hu2024MNRAS}, as a strong magnetic field strength would rather related to the active or flaring state of jet \citep[e.g.,][]{Shukla2020NatCo} than to the average state.

Besides, a smaller $B$ would shift the entire SED towards the longer wavelengths, and a consequent larger $R$ is required to fit a broadband SED.
We have an average $R$ equals to $8.18 \times 10^{16} \, {\rm cm}$, and 18.7\% of sources with $R > 1.0 \times 10^{17} \, {\rm cm}$, which is larger than the typical radius ($< 10^{17} \, {\rm cm}$) of dissipation region.
One need to keep mind that this $R$ should not be seriously considered as the radius of the `one-zone' dissipation region, because we use historical data to form the SED and then fit it with the theoretical model, such a large $R$ should be a cumulative effect.
In other words, during more than 20 years data collecting, the dissipation regions could located in different distance from the central supermassive black hole in different size.
For the parameters of describing the EED, we only set the $\gamma_{\rm min} > 10$ and $\gamma_{\rm max} < 1 \times 10^{7}$.
The average value and corresponding standard deviation of these nine parameters are listed in Table \ref{average_value}.

\begin{table}[htbp]
\centering
\caption{The average value of physical parameters}
\label{average_value}
\begin{tabular}{ll}
\hline
Parameter                   &       Value                   \\
(1)                         &       (2)                     \\
\hline
$\log$R (cm)	     	    &       $16.34 \pm 0.72$        \\
B (Gs)                      &       $0.23 \pm 0.33$         \\
N (cm$^{-3}$)	            &       $608.88 \pm 2296.52$	\\
$\delta$	                &       $20.33 \pm 4.62$        \\
$\log \gamma_{\rm min}$     &       $1.77 \pm 0.33$         \\
$\log \gamma_{\rm 0}$       &       $3.33 \pm 0.31$         \\
$\log \gamma_{\rm max}$     &       $5.17 \pm 0.32$         \\
s				            &       $2.16 \pm 0.13$         \\
r				            &       $0.37 \pm 0.12$         \\
\hline
\end{tabular}
\end{table}

\subsection{The correlation between the shape parameters and implication on Klein-Nishina effect}
In the conventional leptonic one-zone frame of interpreting blazar broadband SED, for the scattering that occurs in the Thomson regime, the SSC peak frequency follows
\begin{equation}
\nu_{\rm ssc} = \frac{4}{3} \gamma_{\rm p}^{2} \nu_{\rm sy},
\label{v_eq}
\end{equation}
where $\gamma_{\rm p}$ stands for the Lorentz factor of electrons that produce most of the synchrotron power and contribute most to the synchrotron peak.
Thus, the $\log \nu_{\rm SSC}$ is naturally linear related to $\log \nu_{\rm sy}$ as $\log \nu_{\rm SSC} = \log \nu_{\rm sy} + const.$, and the constant is determined by $\gamma_{\rm p}$.

\begin{figure}[htbp]
\centering
\includegraphics[scale=0.70]{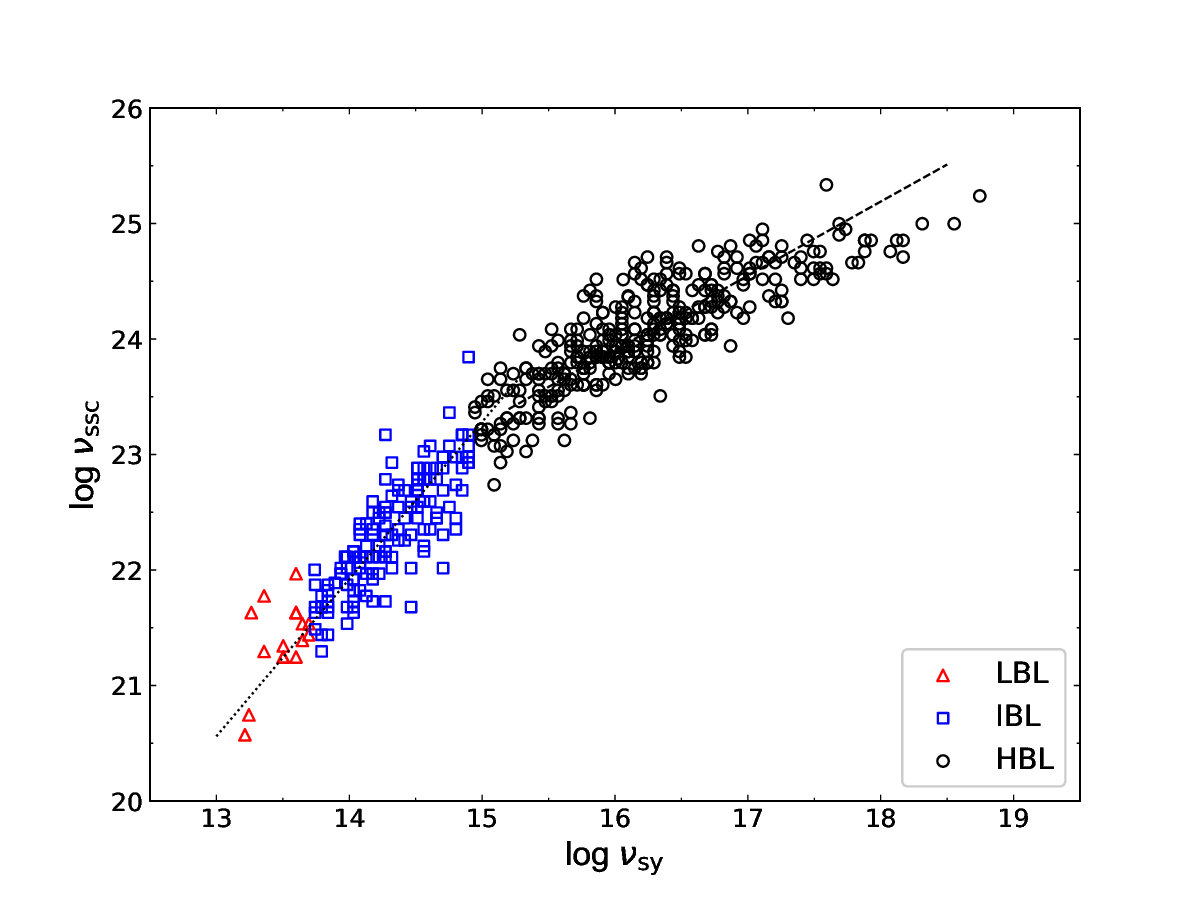}
\caption{The correlations between the synchrotron peak frequency ($\log \nu_{\rm sy}$) and the SSC peak frequency ($\log \nu_{\rm ssc}$).
The red triangle stands for the LBL, the blue square stands for the IBL and the black circle stands for the HBL, the black dotted line represents the linear regression of the combined LBLs and IBLs, and the black dashed line represents the linear regression of the HBLs.}
\label{fre}
\end{figure}

We studied the correlation between the synchrotron peak frequency and the SSC peak frequency, as shown in Figure \ref{fre}.
There is a correlation between the SSC peak frequency and synchrotron peak frequency for the combined LBLs and the IBLs gives
\begin{equation*}
{\rm log} \nu_{\rm ssc} = (1.36 \pm 0.06) {\rm log} \nu_{\rm sy} + (2.87 \pm 0.79),
\end{equation*}
with $q = 0.86$ and $p = 1.0 \times 10^{-48}$;
a correlation between the SSC peak frequency and synchrotron peak frequency for HBLs gives
\begin{equation*}
{\rm log} \nu_{\rm ssc} = (0.64 \pm 0.02) {\rm log} \nu_{\rm sy} + (13.64 \pm 0.33),
\end{equation*}
with $q = 0.85$ and $p = 1.7 \times 10^{-100}$.

For the combined sample of LBLs and IBLs, a slope of $1.36\pm0.06$ is greater than 1, such a slope could caused by several indistinguishable reasons.
It could caused by sample bias, or some other efficient scattering process involved along with the SSC process, or the situation of dissipation region could be more complicated than the typical `one zone' model.

And, we notice a slope of $0.64\pm0.02$, which is significantly smaller than 1, for the HBLs.
This is mainly caused by the Klein-Nishina effect, where the photon energy, measured in the rest frame of the upscattering electron, is comparable to the electron rest-mass energy.
This effect would significantly reduce the cross-section of the scattering and further suppress the IC process, resulting in a relatively smaller $\log \nu_{\rm SSC}$ for given $\log \nu_{\rm sy}$.

Specifically, it is possible to analytically determine the SSC spectrum that tuned by the Klein-Nishina effect and which source could be severely affected through
\begin{equation}
    \gamma_{\rm b}\, \nu{'}_{\rm sy} \geq \frac{3}{4} \frac{m_{\rm e}c^{2}}{h},
\label{KN_condi}
\end{equation}
where the $\gamma_{\rm b}$ is the break Lorentz factor of electrons dominantly produce the emission at the peak frequency, $\nu{'}_{\rm sy}$ is the synchrotron peak frequency in the source frame and related to the observed synchrotron peak frequency as $\nu_{\rm sy} = \nu{'}_{\rm sy} \frac{\delta}{1+z}$, $m_{\rm e}$ is the electron mass, $c$ is the speed of light and the $h$ is the Planck constant \citep{Tavecchio1998}.
Meanwhile, in the leptonic `one-zone' model the $\nu_{\rm sy}$ and $\gamma_{\rm b}$ is correlated through $B$ and $\delta$ as
\begin{equation}
    \nu_{\rm sy} = 3.7 \times 10^{6} \, \gamma_{\rm b}^{2} \, B \frac{\delta}{1+z},
\label{nu_sy}
\end{equation}
note that $\gamma_{\rm b} = \gamma_{\rm 0}$ in this work.
Combining Eq. \ref{KN_condi} and Eq. \ref{nu_sy}, we obtain
\begin{equation}
    \nu_{\rm sy} \geq \nu_{\rm sy}^{\rm c} = 3.17 \times 10^{15} \, B^{1/3} \frac{\delta}{1+z},
\label{nu_sy_KN_condi}
\end{equation}
suggest that source meet this condition should encounter significant Klein-Nishina suppression regarding the SSC process.
We calculated the critical $\nu_{\rm sy}^{\rm c}$ for each source in our sample and found 359 of them show $\nu_{\rm sy} > \nu_{\rm sy}^{\rm c}$.

\subsection{The $\gamma$ ray variability}
Variability is one of the characteristic properties of blazars, have been observed in multi-bands in different timescales.
Generally, variability with timescale of years is classified as long-term one, of days to months is classified as short-term variability \citep{Fan2005ChJAS}.
For the variability show time scale less than one day is called intraday/intranight variability, or micro-variability \citep{Singh2020AN, Amaya2022ApJ, Otero2022MNRAS}.
The idea of explaining variability relates to orbiting disturbance, jet spiral structure, precession, geometric effect, etc \citep{Camenzind1992AA, Gopal1992AA}.
The fastest variability at time scale of minutes is mostly observed in the high energy band, particularly in the $\gamma$-ray band \citep{Aharonian2007ApJ, Aleksic2011, Ackermann2016ApJL}, and are used to constrain the emission region size according to the reason of causality.
However, the stochastic nature of the blazar variability is not fully understood.
In our previous work, \citet{Fan2023ApJS} explored the reason of raising variability which is the initial seed instability, and suggested that the accumulated instability demonstrated as variability.
We also found that Fermi blazars could show the Kelvin–Helmholtz instability \citep{Sol1989MNRAS} in the emission zone, the instability could occur when the magnetic field is weaker than the critical magnetic field, the critical magnetic field strength ($B_{\rm c}$) is mainly proportional to the square root of the particle number density.
In other words, the instability and thus the variability may be compressed if $B$ is too strong.

\begin{figure}[htbp]
\centering
\includegraphics[scale=0.7]{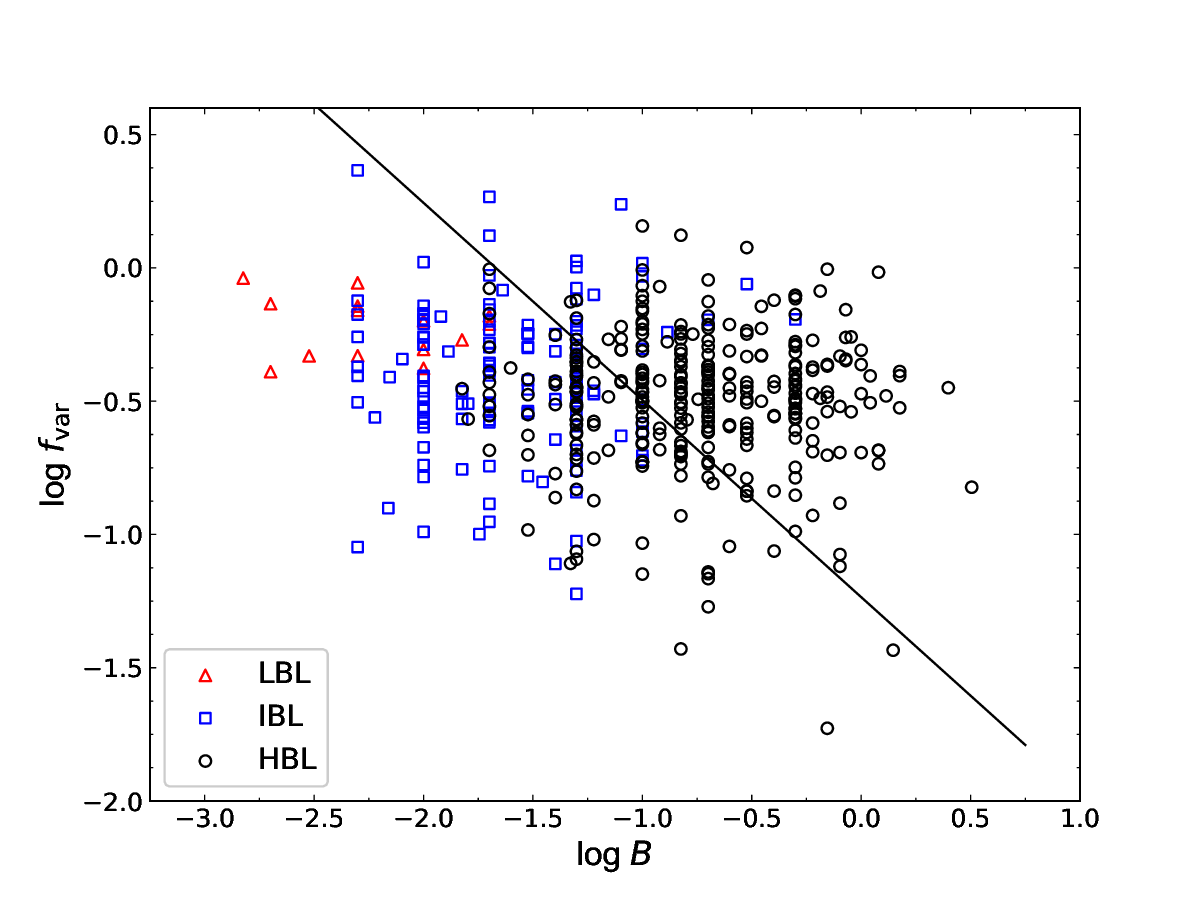}
\caption{The correlations between the magnetic field strength ($\log B$) and the $\gamma$-ray fractional variability ($\log f_{\rm var}$).
The red triangle stands for the LBL, the blue square stands for the IBL and the black circle stands for the HBL.
The solid black line represents the linear regression result of these BL Lacs.}
\label{bfv}
\end{figure}

In this work, we studied the correlation between the magnetic field strength and the $\gamma$-ray band fractional variability obtained from 4FGL \citep{Abdollahi2020}, and the results are shown in Figure \ref{bfv} and given in 
\begin{equation*}
{\rm log} f_{\rm var} = -(0.74 \pm 0.06) {\rm log} B - (1.24 \pm 0.06),
\end{equation*}
with $q = -0.17$ and $p = 3.0 \times 10^{-4}$;
The anti-correlation between $\log f_{\rm var}$ and $\log B$ is very weak, as shown in Figure \ref{bfv}, and could be influenced by a few outliers, which are located in the lower-right region of the Figure \ref{bfv}.
The `column' clustering data of $\log B$, that caused by our semi-manual SED modeling approach, would not affect the anti-correlation significantly, a Gaussian perturbation is employed to validate the correlation (see in Appendix B).
It seems that our result is consistent with the prediction that the magnetic field suppresses the development of instabilities. 
However, this conclusion should be treated with caution and requires further verification in future studies.

\subsection{Particle Acceleration Mechanism}
The particle acceleration mechanism is key to understand how the particles are accelerated and propagating in the universe.
It has been studied and discussed by many authors \citep{Ball1992, Blandford1994, Katarzynski2006, Petrosian2008, Tramacere2011}.
In this work, we applied the `LPPL' function for the EED during modelling the SEDs.
Several mechanisms can lead to the electron energy distribution following a log-parabola function through statistical processes, 
including energy-dependent probability acceleration \citep[EDPA,][]{Massaro2004a}, 
fluctuation on fractional energy gain acceleration in the frame of stochastic acceleration \citep[FFGA,][]{Tramacere2011} 
and `quasi-'monoenergetic particle injection (QMPI) in the frame of stochastic acceleration.

\citet{Massaro2004a} presented some phenomenological considerations on the particle spectra arising from statistical acceleration.
They showed that it is possible to obtain an integral energy distribution for the particles that follows a log-parabolic law under quite reasonable hypotheses, such as the energy dependence of both the fractional acceleration gain and the acceleration probability.
Thus, the spectral curvature $b_{\rm sy}$ of a synchrotron emission spectrum is expected to be correlated with the electron spectrum curvature $r$ as $b_{\rm sy}=r/4$.
Under this process, \citet{Chen2014} suggested the the synchrotron peak frequency $\log \nu_{\rm sy}$ is correlated with $b_{\rm sy}$, in the form of
\begin{equation}
\log \nu_{\rm sy} \approx 2/(5b_{\rm sy}) + m
\end{equation}
with the condition of an electron acceleration efficiency inversely proportional to energy while the fractional energy gain stays constant.

\citet{Tramacere2011} gave a statistical description for the case of fluctuation on fractional energy gain, which is a stochastic process. 
They obtained a log-parabola law for the electron energy distribution.
\citet{Chen2014} suggested that in the case of fluctuation on fractional energy gain, the synchrotron peak frequency and the synchrotron curvature should give
\begin{equation}
\log \nu_{\rm sy} \approx 3/(10b_{\rm sy}) + m.
\end{equation}

In the scenario of the stochastic mechanism, using the Fokker-Planck equation with the presence of a momentum-diffusion term, the log-parabolic distribution can also be derived from a `quasi-' mono-energetic and instantaneous injection \citep{Kardashev1962, Tramacere2011} and a correlation of
\begin{equation}
\log \nu_{\rm sy} \approx 1/(2b_{\rm sy}) + m
\end{equation}
was suggested by \citep{Chen2014}.

\begin{figure}[htbp]
\centering
\includegraphics[scale=0.7]{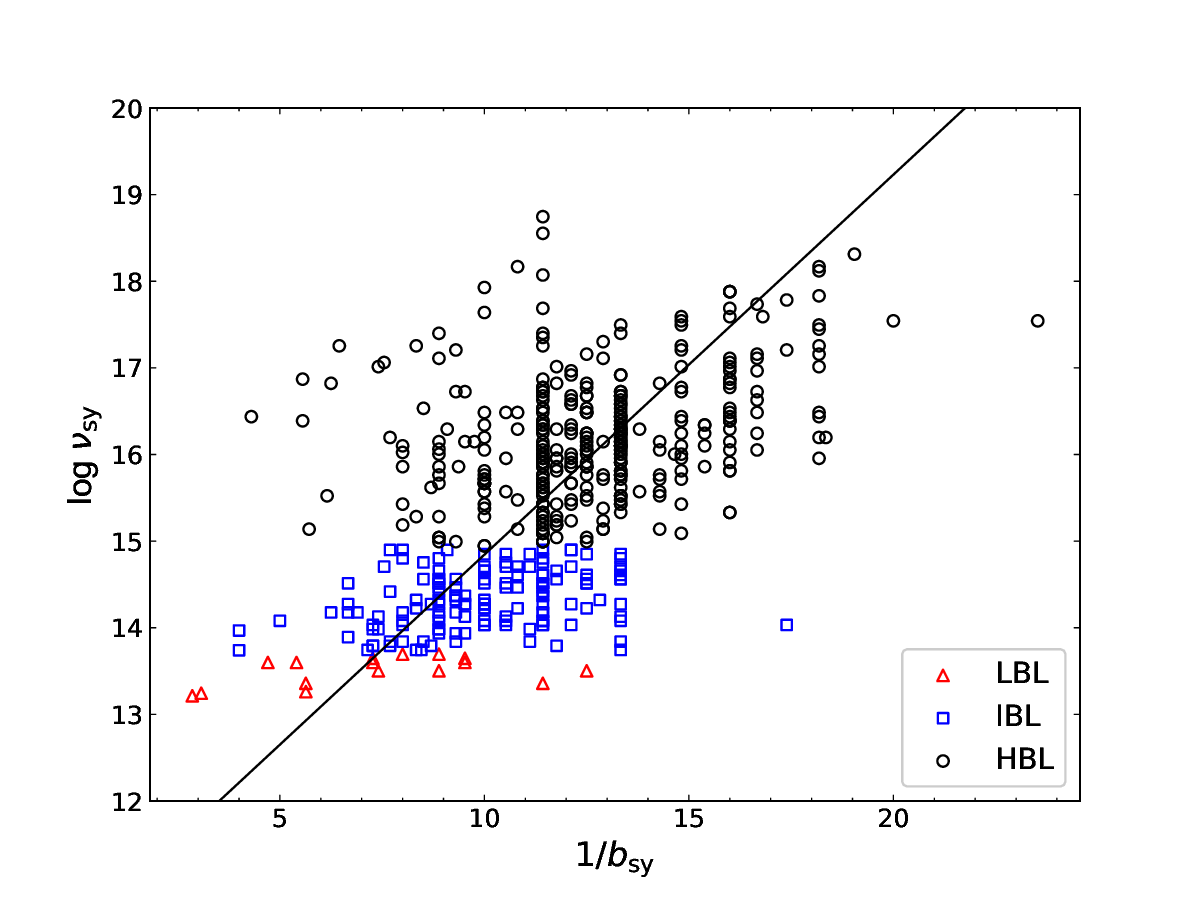}
\caption{The correlation between synchrotron peak frequency ($\log \nu_{\rm sy}$) and the inverse synchrotron spectral curvature ($1/b_{\rm sy}$).
The red triangle stands for the LBL, the blue square stands for the IBL and the black circle stands for the HBL, the solid black line represents the linear regression result of these BL Lacs.}
\label{fre_b}
\end{figure}

In order to study the particle acceleration of blazars in different types, we plot $1/b_{\rm sy}$ vs. $\log \nu_{\rm sy}$ in Figure \ref{fre_b} for BL Lacs in this work.
In the case of all BL Lacs in our sample, the linear regression gives
\begin{equation*}
\log \nu_{\rm sy} = (0.44 \pm 0.02) (1/b_{\rm sy}) + (10.46 \pm 0.22),
\end{equation*}
with a $q = 0.56$ and $p = 1.2 \times 10^{-43}$;
The slope of correlation for the entire sample of BL Lacs in this work gives $k_{\rm all} = 0.44 \pm 0.02$, which can be explained by a EDPA mechanism.
This result is consistent with the results in our previous work \citet{Xiao2024ApJ966}, in which we studied the particle acceleration problem with a larger blazar sample and the result suggested that those high inverse Compton peak sources (HCPs, mostly BL Lacs) follow the acceleration mechanism of EDPA, and such a acceleration mechanism is believed to accelerate particles more efficiently in case of BL Lacs.

\subsection{The estimation of jet power}
Jet power, $P_{\rm jet}$, is a crucial parameter for studying and understanding blazar jets' composition, launching, radiation and acceleration.
There are several methods have been proposed to estimate jet power, such as the one-zone leptonic model \citep{Ghisellini2009MNRAS397}, radio core shifts \citep{Zdziarski2015MNRAS451}, extended radio emission \citep{Willott1999MNRAS309}, high-energy gamma-ray luminosity \citep{Ghisellini2014, Nemmen2012}, and model-based radio emission \citep{Foschini2024Universe}.

Based on the leptonic one-zone model, we calculate the jet power from relativistic electrons ($P^{\rm jet}_{\rm e}$), magnetic field (or Poynting flux, $P^{\rm jet}_{\rm B}$), radiation ($P^{\rm jet}_{\rm rad}$) and cold proton kinetic ($P^{\rm jet}_{\rm p}$) \citep{Celotti2008, Ghisellini2010, Xiao2024RAA24} as:
\begin{equation}
P^{\rm jet}_{\rm tot} = \Sigma_{i} P^{\rm jet}_{i} = \Sigma_{i} \pi R_{\rm diss}^{2} \Gamma^{2} c U_{i}, 
\label{Pjet}
\end{equation}
where $U_{i}$ stands for the energy density of the magnetic field ($i=B$), the relativistic electron ($i=e$), cold proton ($i=p$), and bolometric radiation ($i=r$) in the comoving frame, and can be further expressed as:
\begin{equation}
U_{\rm e} = m_{\rm e}c^{2} \int N_{\rm e}(\gamma)\gamma d\gamma, 
\label{Ue}
\end{equation}
\begin{equation}
U_{\rm p} = m_{\rm p}c^{2} \int N_{\rm p}(\gamma) d\gamma,
\label{Up}
\end{equation}
\begin{equation}
U_{\rm B} = B^{2}/8\pi, 
\label{UB}
\end{equation}
\begin{equation}
U_{\rm rad} = L_{\rm obs}/(4\pi R^{2}c\delta^{2}), 
\label{Ur}
\end{equation}
where $L_{\rm obs}$ is the observed nonthermal bolometric jet luminosity.
Assuming the jet charge neutrality holds for bazar jets and the neutrality is maintained through electrons and protons, in this work, we employ the ratio of the number of cold proton to the number of relativistic electron is one ($N_{\rm p} = N_{\rm e}$).
These calculated results of jet powers are listed in Table \ref{power}.

The $\gamma$-ray luminosity can be calculated as
\begin{equation}
L_{\rm \gamma} = 4\pi d_{\rm L}^2(1+z)^{(\alpha_{\gamma}-2)}F,
\end{equation}
where $d_{\rm L} = \frac{c}{H_{\rm 0}}\int^{1+z}_{1}\frac{1}{\sqrt{\Omega_{\rm m}x^{3}+1-\Omega_{\rm m}}}dx$ is a luminosity distance \citep{Komatsu2011}, $F$ is the $\gamma$-ray band flux that can be obtained through `Flux1000' (the integral photon flux from 1 to 100 GeV) and $\alpha_{\gamma}$, which is the photon index, and $(1+z)^{(\alpha_{\gamma}-2)}$ stands for a $K$-correction.

Recently, \citet{Foschini2024Universe} presented a method of estimating blazar jet power based on radio observations and the model proposed by \citet{Blandford1979}, the total jet power can be estimated by
\begin{equation}
P^{\rm jet}_{\rm radio} = (4.5 \times 10^{44}) \lgroup \frac{S_{\rm \nu} d^{2}_{\rm L,9}}{1+z} \rgroup^{\frac{12}{17}}, 
\label{Pradio}
\end{equation}
where $S_{\nu}$ is the observed radio flux density in units of Jy, $d_{\rm L,9}$ is the luminosity distance in units of Gpc.

\begin{figure}[htbp]
\centering
\includegraphics[scale=0.7]{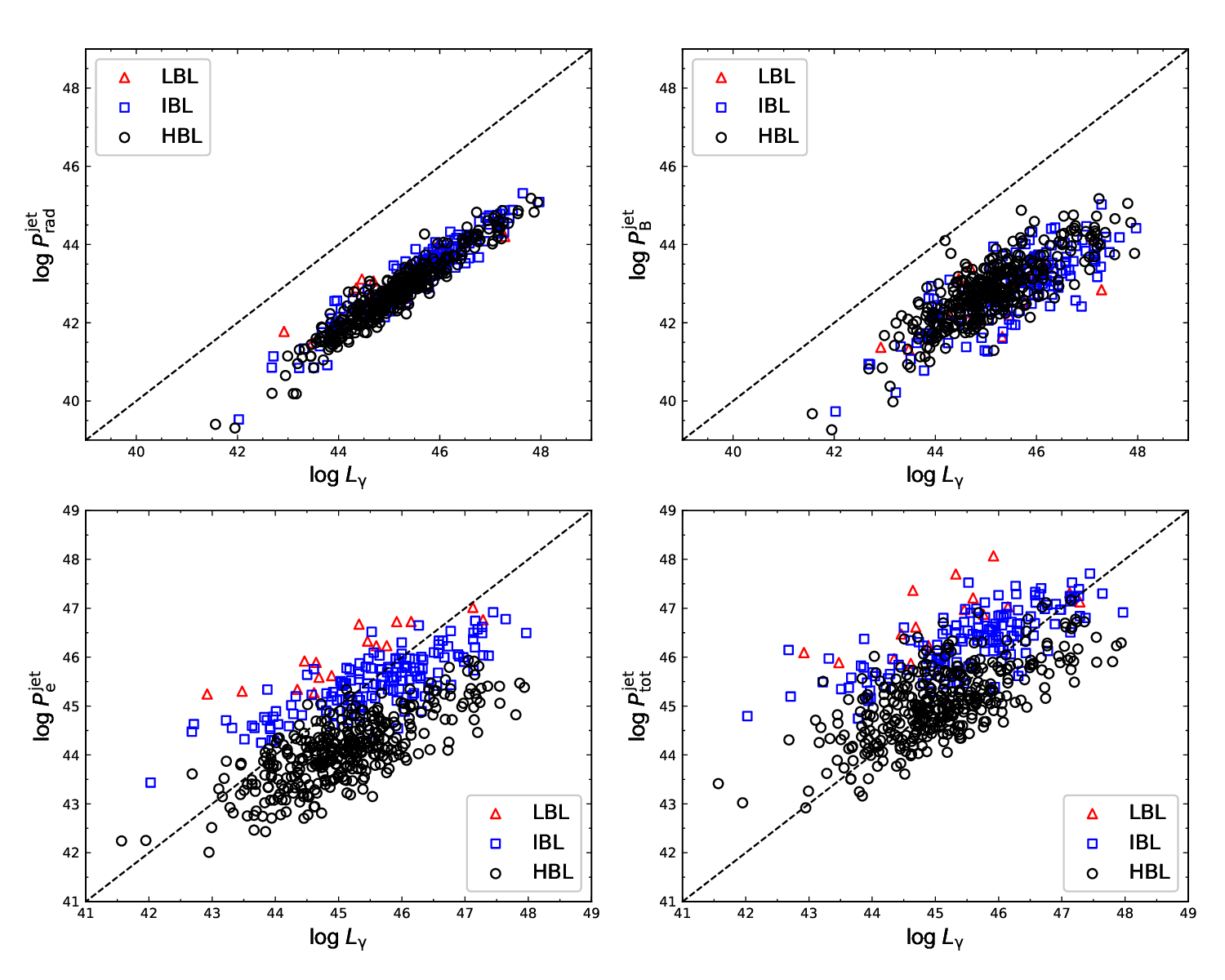}
\caption{The comparison between SED jet power and gamma-ray luminosity.
The symbols are defined the same as previous, and the dashed line represents the equality line.}
\label{Pjet_Lgamma}
\end{figure}

\begin{figure}[htbp]
\centering
\includegraphics[scale=0.7]{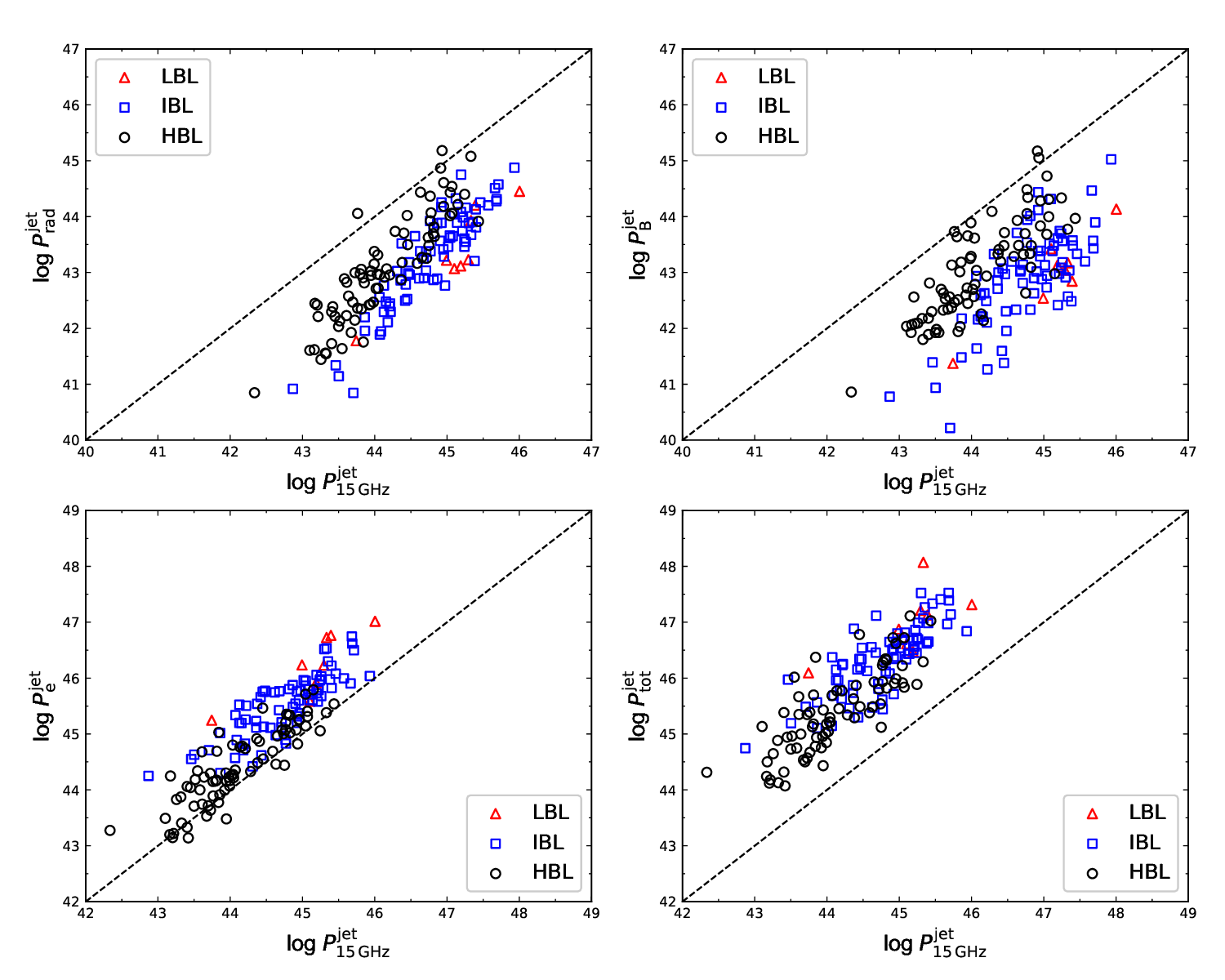}
\caption{The comparison between SED jet power and  radio jet power.
The symbols are defined the same as previous, and the dashed line represents the equality line.}
\label{Pjet_Pradio}
\end{figure}

We calculate $P^{\rm jet}_{\rm e}$, $P^{\rm jet}_{\rm B}$, $P^{\rm jet}_{\rm rad}$, $P^{\rm jet}_{\rm tot}$, and $L_{\rm \gamma}$ for the entire sample of bright \textit{Fermi} BL Lacs.
Meanwhile, we manage to collect 15 GHz radio data from literature for 159 sources, and the averaged flux density is used to calculate $P^{\rm jet}_{\rm 15 \, GHz}$.
The comparisons between the one-zone leptonic model jet power and the $\gamma$-ray luminosity is shown in Figure \ref{Pjet_Lgamma}, and the radio jet power is shown in Figure \ref{Pjet_Pradio}.

Our results suggest that the $\gamma$-ray luminosity is a good representative of the entire jet power of the bright \textit{Fermi} BL Lacs, while, the $P_{\rm tot}^{\rm jet}$ is slightly larger than the $L_{\rm \gamma}$ for LBLs and IBLs, 
the $P_{\rm tot}^{\rm jet}$ is compatible to $L_{\rm \gamma}$ for HBLs
as shown in the lower-right panel of Figure \ref{Pjet_Lgamma}.
The total jet power $P_{\rm tot}^{jet}$ is dominated by the kinetic power of relativistic electrons and cold protons ($P_{\rm tot}^{\rm jet}$, which is not shown in Figure \ref{Pjet_Lgamma}), and the $L_{\rm \gamma}$ is significantly larger than $P_{\rm B}$ and $P_{\rm rad}$ as shown in the upper panels of Figure \ref{Pjet_Lgamma}.
However, the method of estimating jet power with 15 GHz radio observation is not consistent with results of the jet power that we obtained via SED modelling. 
Specifically, $P_{15 \, GHz}^{\rm jet}$ is significantly greater than both $P^{\rm jet}_{\rm B}$ and $P^{\rm jet}_{\rm rad}$, and it is significantly smaller both $P^{\rm jet}_{\rm rad}$ and $P^{\rm jet}_{\rm tot}$, as shown in Figure \ref{Pjet_Pradio}.

\begin{table}[htbp]
\scriptsize
\centering
\caption{The estimation of jet power}
\label{power}
\begin{tabular}{lccccccccc}
\hline
4FGL name & $P^{\rm jet}_{\rm rad}$  &  $P^{\rm jet}_{\rm B}$ & $P^{\rm jet}_{\rm e}$ & $P^{\rm jet}_{\rm p, cold}$ & $P^{\rm jet}_{\rm tot}$ & Flux1000 & $\alpha_{\gamma}$ & $S_{\rm 15 GHz}$ & Ref.  \\
    &  ${\rm erg/s}$ &  ${\rm erg/s}$ &   ${\rm erg/s}$  & ${\rm erg/s}$ &  ${\rm erg/s}$  &  ${\rm photon/cm^{2}/s}$   &     &  ${\rm Jy}$   &       \\
(1) & (2) & (3) & (4) &    (5)          & (6) & (7) & (8) & (9) & (10)  \\
\hline
J0001.2-0747 &	3.01E+42 &	1.62E+42 &	1.57E+45 &	7.64E+45 &	9.22E+45 &	7.02E-10 &	2.08 &	0.169	            &	R14	\\
J0003.9-1149 &	1.61E+43 &	3.06E+42 &	9.50E+45 &	3.24E+46 &	4.19E+46 &	3.23E-10 &	2.04 &	0.716, 0.618	    &	R11, R14	\\
J0006.3-0620 &	1.31E+43 &	1.35E+43 &	8.32E+45 &	2.14E+46 &	2.97E+46 &	1.21E-10 &	2.17 &	2.283, 2.41, 2.198	&	L11, R11, R14	\\
J0009.1+0628 &	1.35E+44 &	3.74E+43 &	1.67E+46 &	7.98E+46 &	9.66E+46 &	5.10E-10 &	2.22 &	0.223	            &	R14	\\
J0009.8-4317 &	4.77E+42 &	6.61E+42 &	9.08E+43 &	1.87E+45 &	1.97E+45 &	2.35E-10 &	2.12 &		&		\\
...          &  ...      & ...       & ...       &  ...      &  ...      &  ...      &  ...  &  ... &   ...  \\
\hline
\end{tabular}
\tablecomments{
Column (1) 4FGL name;
column (2) the jet radiation power;
column (3) the jet power carried by magnetic field;
column (4) the jet power carried by electrons;
column (5) the jet power carried by cold protons;
column (6) the total jet power;
column (7) the integral photon flux from 1 to 100 GeV;
column (8) the photon index of `Flux1000';
column (9) the radio flux density;
column (10) reference to the $S_{\rm 15GHz}$, 
`D78' stands for \citet{Dell1978ApJ224},
`A10' stands for \citet{Abdo2010ApJ716}, 
`A11' stands for \citet{Ackermann2011ApJ741}, 
`L11' stands for \citet{Lister2011ApJ742}, 
`R11' stands for \citet{Richards2011ApJS194}, 
`R14' stands for \citet{Richards2014MNRAS438}.
Only five items are displayed, the entire table is available in machine-readable form.}
\end{table}

\section{Summary and Conclusion}\label{sec: con}
For the purpose of investigating the general property of blazars, for the first time, we modelled broadband (radio to high-energy $\gamma$-ray) average-state SEDs for 513 bright \textit{Fermi} BL Lac objects using archival data through physics-determined SED method.
Our main results and conclusion are as follows:
\begin{itemize}
\item [1.] 
We have modelled broadband average-state SED for 513 \textit{Fermi} bright BL Lacs, and obtained shape parameters of the SED ($\log \nu_{\rm sy}$, $\log \nu_{\rm sy}f_{\rm sy}$, $\log \nu_{\rm ssc}$, and $\log \nu_{\rm ssc}f_{\rm ssc}$) and physical parameters of describing the dissipation region and the EED ($B$, $R_{\rm diss}$, $\delta$, $N$, $\gamma_{\rm min}$, $\gamma_{\rm 0}$, $\gamma_{\rm max}$, $s$ and $r$);

\item [2.]
Through comparison, we found that the physics-determined SED show significant differences in shape parameters with respect to that obtained via the statistic-determined SED.
The differences on $\log \nu_{\rm sy}$ would lead to misclassification and the overestimation of $\log (\nu_{\rm sy}f_{\rm sy})$ could result from a disk thermal emission;
Besides, we have relatively weaker $B$ and larger $R$ than those obtained using the simultaneously observed data, suggesting the dissipation region is more extended and less magnetized.

\item [3.]
The correlation between $\log \nu_{\rm ssc}$ and $\log \nu_{\rm sy}$ reveals a slope of 1.36 for the combined LBLs and IBLs, suggests the IC process in these sources could be more complex; a slope of 0.64 for the HBLs suggests a significant Klein-Nishina suppression effect.
A calculation of $\nu_{\rm sy}^{\rm c}$ suggests 359 out of 513 sources should suffer Klein-Nishina supression;


\item [4.]
The correlations between synchrotron curvature and peak frequency gave $\log \nu_{\rm sy} \propto 0.44 \frac{1}{b_{\rm sy}}$ for the entire sample of this work, and suggests a mechanism of EDPA dominating the particle acceleration in jets;

\item [5.]
A comparison between SED jet power and the $\gamma$-ray luminosity, and the radio estimated jet power, suggests that the $\gamma$-ray luminosity is a better estimator than 15 GHz radio luminosity to estimate the total jet power.
\end{itemize}

\acknowledgments
H.B.X acknowledges the support from the National Natural Science Foundation of China (NSFC 12203034), the Shanghai Science and Technology Fund (22YF1431500), the science research grants from the China Manned Space Project (CMS-CSST-2025-A07), and the Shanghai Municipal Education Commission regarding artificial intelligence empowered research.
R.X acknowledges the support from the NSFC under grant No. 12203043.
J.H.F acknowledges the support from the NSFC U2031201, NSFC 11733001, NSFC 12433004, the Scientific and Technological Cooperation Projects (2020–2023) between the People’s Republic of China and the Republic of Bulgaria, the science research grants from the China Manned Space Project with No. CMS-CSST-2021-A06, and the support for Astrophysics Key Subjects of Guangdong Province and Guangzhou City.
This research was partially supported by the Bulgarian National Science Fund of the Ministry of Education and Science under grants KP-06-H38/4 (2019), KP-06-KITAJ/2 (2020) and KP-06-H68/4 (2022).
S.H.Z acknowledges support from the National Natural Science Foundation of China (Grant No. 12173026), the National Key Research and Development Program of China (Grant No. 2022YFC2807303), the Shanghai Science and Technology Fund (Grant No. 23010503900), the Program for Professor of Special Appointment (Eastern Scholar) at Shanghai Institutions of Higher Learning and the Shuguang Program (23SG39) of the Shanghai Education Development Foundation and Shanghai Municipal Education Commission.

\appendix
\section{The distribution of SED physical parameter}\label{app_a}
\begin{figure}[htbp]
\centering
\includegraphics[scale=0.55]{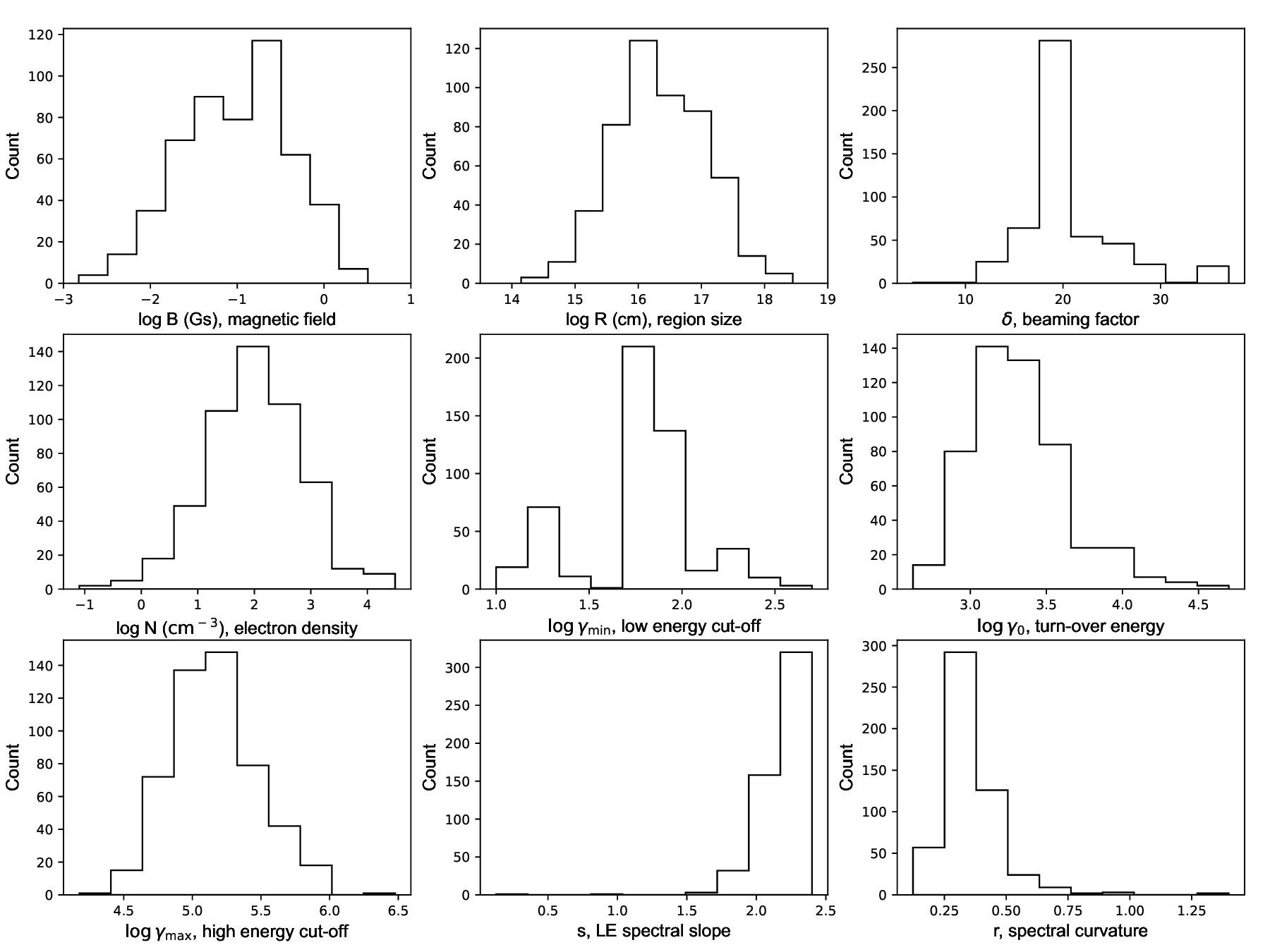}
\caption{The distribution of SED physical parameters.}
\label{Para_hist}
\end{figure}

We illustrate the distribution of the physical parameters, as shown in Figure \ref{Para_hist}, that obtained through the broadband SED fitting in this work.

\section{Perturbation for magnetic field strength and spectral curvature}\label{app_b}
\begin{figure}[htbp]
\centering
\includegraphics[scale=0.7]{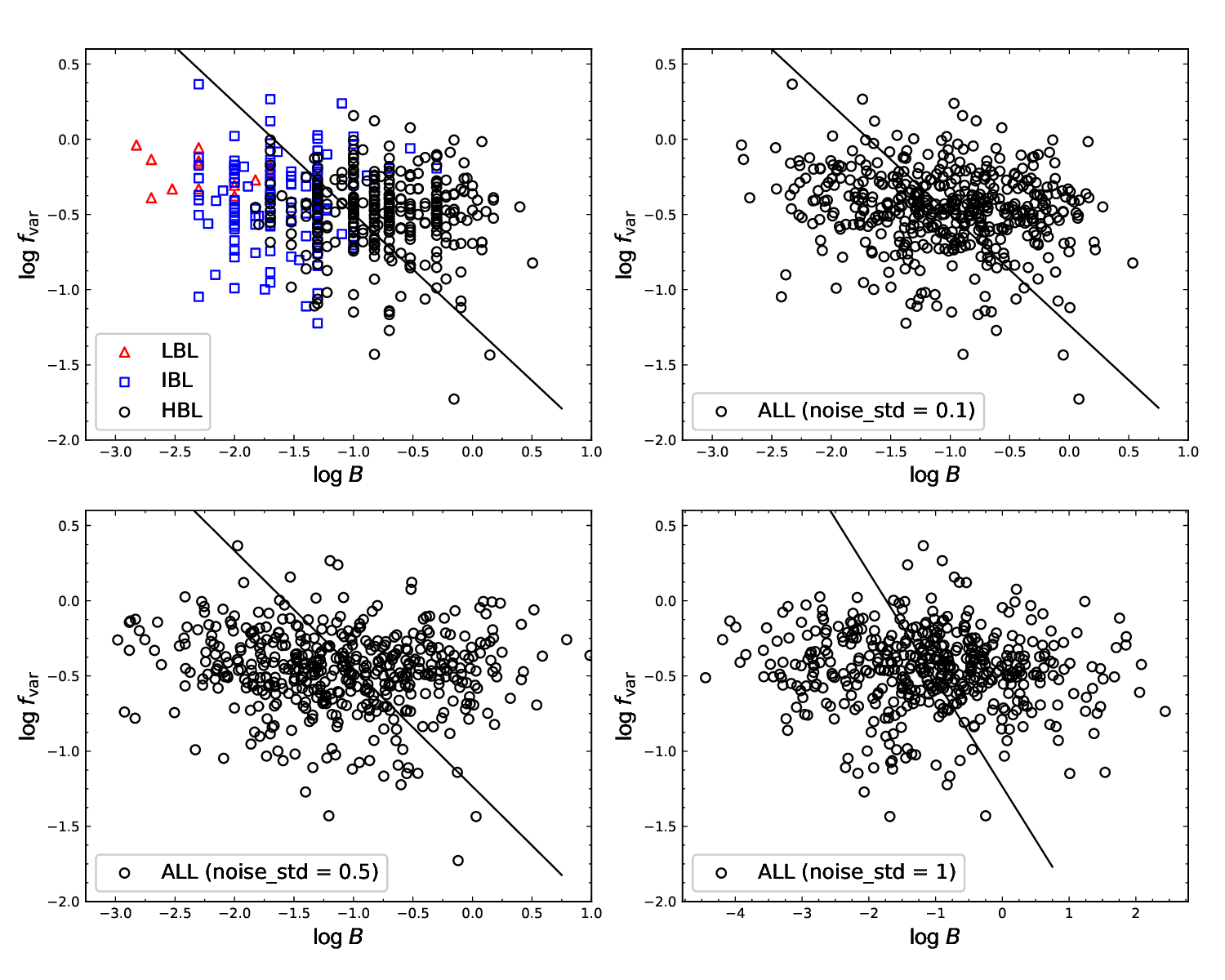}
\caption{The correlations between the perturbed magnetic field strength ($\log B$) and the $\gamma$-ray fractional variability ($\log f_{\rm var}$).
The upper-left panel gives the original correlation as we have shown in Figure \ref{bfv}, and the rest of the three panels show the Gaussian perturbed results.}
\label{bfv_com}
\end{figure}

\begin{figure}[htbp]
\centering
\includegraphics[scale=0.7]{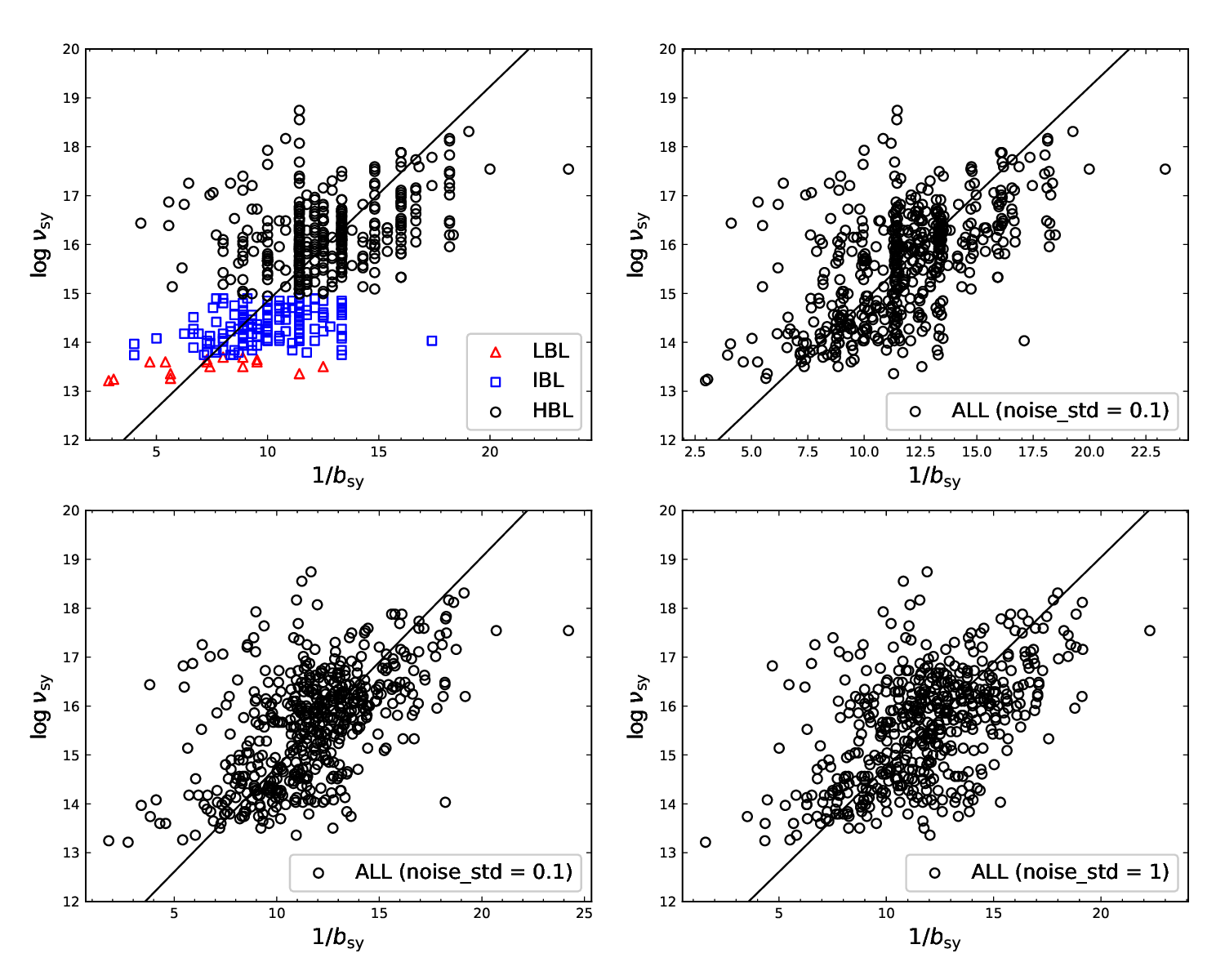}
\caption{The correlation between synchrotron peak frequency ($\log \nu_{\rm sy}$) and the perturbed inverse synchrotron spectral curvature ($1/b_{\rm sy}$).
The upper-left panel gives the original correlation as we have shown in Figure \ref{fre_b}, and the rest of the three panels show the Gaussian perturbed results.}
\label{fre_b_compare}
\end{figure}

For the purpose of validating the anti-correlation between $\log f_{\rm var}$ and $\log B$, and the positive correlation between $\log \nu_{\rm sy}$ and $1/b_{\rm sy}$, we employ Gaussian perturbation.
Specifically, we add zero-mean, standard deviation Gaussian (normal) noise to the original variable $\log B$ and $b_{\rm sy}$, and three noise ratio (0.1, 0.5, 1.0) are used to test the validation.
The results are shown in Figure \ref{bfv_com}, Figure \ref{fre_b_compare}, and Table \ref{com_tab}, these results suggest that the anti-correlation between $\log f_{\rm var}$ and $\log B$ for both `nosie\_std=0.1' and `noise\_std=0.5' and the positive correlation between $\log \nu_{\rm sy}$ and $1/b_{\rm sy}$ for all three cases of noise\_std. 

\begin{table}[htbp]
\centering
\caption{The linear regression results with perturbed parameters.}
\label{com_tab}
\begin{tabular}{lcccccc}
\hline
Correlation & Noise ratio & N & $\beta$ & $\alpha$ & q & p  \\
(1) & (2) & (3) & (4) & (5) & (6) & (7) \\
\hline
                                      & noise\_std = 0.1	& 466	& $-0.73 \pm 0.06$	& $-1.23 \pm 0.06$	  & 0.17  & $2.4 \times 10^{-4}$    \\
$\log f_{\rm var}$ vs $\log B$        & noise\_std = 0.5  & 466	  & $-0.78 \pm 0.08$  & $-1.33 \pm 0.09$	& 0.10  & 0.03   \\
                                      & noise\_std = 1.0	& 466	& $-0.71 \pm 0.12$	& $-1.22 \pm 0.13$	  & 0.08  & 0.08   \\ \hline
                                      & noise\_std = 0.1	& 513	& $0.44 \pm 0.02$	& $10.46 \pm 0.22$	  & 0.56  & $6.1 \times 10^{-44}$   \\
$\log \nu_{\rm sy}$ vs $1/b_{\rm sy}$ & noise\_std = 0.5  & 513	  & $0.43 \pm 0.02$	  & $10.57 \pm 0.23$    & 0.55  & $1.7 \times 10^{-41}$   \\
                                      & noise\_std = 1.0	& 513	& $0.43 \pm 0.02$	& $10.54 \pm 0.22$	  & 0.53  & $1.7 \times 10^{-38}$   \\
\hline
\end{tabular}
\tablecomments{
column (1) gives the name of correlation;
column (2) gives the noise ratio;
column (3) is the number of the sources;
column (4) gives the slope;
column (5) gives the intercept;
column (6) gives the Pearson correlation coefficient;
column (7) gives the chance probability;
}
\end{table}

\bibliography{lib_xiao}{}
\bibliographystyle{aasjournal}

\end{document}